\documentclass{pasa}%
\pdfoutput=1

\usepackage{graphicx}

\title[Clusters towards W\,31 complex]{VVV analysis of star clusters towards the W\,31 star-forming complex
}

\author[Bianchin et al.]{M. Bianchin$^1$, E. F. Lima$^2$, E. Bica$^3$, R.~A. Riffel$^{1,4}$, C. Bonatto$^3$ and R.~K. Saito$^{5}$
\affil{$^1$Departamento de F\'\i sica, CCNE, Universidade Federal de Santa Maria,  97105-900, Santa Maria, RS, Brazil\\
$^2$Universidade Federal do Pampa, Campus Uruguaiana, CP 118, 97508-000, Uruguaiana, RS, Brazil\\
$^3$Departamento de Astronomia, IF, Universidade Federal do Rio Grande do Sul, CP 15051, 91501-970, Porto Alegre, RS, Brazil\\
$^4$Department of Physics and Astronomy, Johns Hopkins University, Baltimore, MD 21218, USA \\
$^5$Departamento de F\'\i sica, Universidade Federal de Santa Catarina, 88040-900, Florian\'opolis, SC, Brazil
}
}%

\jid{PASA}
\doi{10.1017/pas.\the\year.xxx}
\jyear{\the\year}

\usepackage{aas_macros}
\usepackage{hyperref} 
\hypersetup{colorlinks,citecolor=blue,linkcolor=blue,urlcolor=blue}

\begin{document}

\begin{frontmatter}
\maketitle

\begin{abstract}
The giant H{\sc ii} region W\,31 hosts the populous star cluster W\,31-CL and others projected on or in the surroundings. The most intriguing object is the stellar cluster SGR\,$1806-20$, which appears to be related to a Luminous Blue Variable (LBV) -- a luminous supergiant star.  We used the deep VVV {\it J-}, {\it H-} and {\it K$_S$}-bands photometry combined with 2MASS data in order to address the distance and other physical and structural properties of the clusters W\,31-CL, BDS\,113 and SGR\,$1806-20$. Field-decontaminated photometry was used to analyse colour-magnitude diagrams and stellar radial density profiles, using procedures that our group has developed and employed in previous studies.  We conclude that the clusters W\,31-CL and BDS\,113 are located at $4.5$\,kpc and $4.8$\,kpc and have ages of  $0.5$\,Myr and $1$\,Myr, respectively. This result, together with the pre-main sequence (PMS) distribution in the colour-magnitude diagram, characterises them as members of the W\,31 complex. The present photometry detects the stellar content, addressed in previous spectroscopic classifications, in the direction of  the cluster SGR\,$1806-20$, 
including the LBV, WRs, and  foreground stars. We derive  an age of $10\pm4$\,Myr and a distance of {$d_{\odot}=8.0\pm1.95$\,kpc}. The cluster is extremely absorbed, with $A_V=25$\,mag.  The present results indicate that SGR\,$1806-20$ is more distant by a factor  1.8 with respect to the W\,31 complex, and thus not physically  related to it.

\end{abstract}
\end{frontmatter}

\section{INTRODUCTION }
\label{intro}
Young stellar clusters are the ideal laboratories for understanding star formation processes.  Star forming regions originate from Giant Molecular Clouds (GMC) and in such scenario many different physical phenomena can occur, such as H{\sc ii} regions, stellar evolution, embedded clusters (ECs) and SNe. \citet{lada03} argue that the association between the remaining GMC gas, or H{\sc ii} region, and the newborn stellar clusters do not last longer than 5\,Myr. The W\,31 region hosts two stellar clusters that are examples of these associations \citep{beuther11}. {Besides, it might be physically  connected to a Soft Gamma-Ray Repeater (SGR) \citep[and references therein]{bibby08}, one among the only four observed to date. }

The W\,31 complex \citep{westerhout58} is a giant star forming region in the Milky Way \citep{blum01} with a luminosity of $L\sim 6\times 10^6$\,L$_{\odot}$, by assuming a distance of 6\,kpc \citep{kim02}. {According to \citet{beuther11}, W\,31 is composed of two H{\sc ii} regions, two ultra-compact H{\sc ii} (UCH{\sc ii})} regions and at least one starless clump, indicating different star formation episodes. \citet{wilson72} defines the W\,31 complex as the association among three H{\sc ii} regions. Two of them are the same as \citet{beuther11} and the other is G\,$10.6-0.4$. The  two H{\sc ii} regions studied by \citet{beuther11} are associated with the infrared clusters W\,31-CL \citep{blum01} and BDS\,113 \citep{bica03}. The stellar content of the W\,31 complex has been addressed in two ways in previous studies. \citet{blum01} studied a $4' \times 4'$ region encompassing the central cluster W\,31-CL. Their analysis differs from the present one mostly in the sense that we include the effects of the dense background stellar field. {\citet{dewangan15} used star counts (including very faint PMS stars) to study the infrared bubble CN\,148 and found several sub-clustering, while in the present study we analyse the cluster BDS\,113, that is a prominent cluster with a core in W\,31. The complex also includes the cluster candidate or loose embedded group BDS 112 \citep{bica03}, which analysis is beyond the scope of the present study.}


Projected {near the confirmed W\,31 stellar and gaseous components }lies SGR\,$1806-20$ \citep{kulkarni93}.  This object is associated with a supernova remnant (G\,$10.0-0.3$), an X-ray source AX\,$1805.7-2025$ \citep{murakami94},  and an infrared stellar cluster \citep{fuchs99, eikenberry04}. The stellar cluster is a matter of study itself, thus it contains a candidate Luminous Blue Variable (LBV) \citep{kulkarni95}, three Wolf-Rayet (WR) and one OB supergiant \citep{figer05}.  { \citet{eikenberry04} obtained a distance of $d_{\odot}\sim15$\,kpc for SGR\,1806$-$20, which was called the far component of W\,31 complex.}


In order to overcome the extinction effects we perform a near infrared (NIR) photometric study of the W\,31 complex and SGR\,$1806-20$. The  data come from two different surveys, {\em VISTA Variables in the V\'{\i}a L\'actea} (VVV, \citealp{minniti10}) and the {\em Two Micron All Sky Survey} (2MASS) \citep{Skrutskie}. {The former provides a limiting magnitude $J\approx 17$\,mag and the photometry with an adequate S/N throughout. 2MASS photometry is used to overcome the saturation effect in the bright end.}
{We used a field-star (FS) decontamination routine \citep{bonatto07} to analyse the cluster colour-magnitude diagrams (CMDs). This tool is very effective in such dense stellar fields. The age, distance and extinction were determined by the PARSEC isochrones \citep{bressan12} fitted to the CMDs. We also studied the cluster structure through the radial density profiles (RDPs). Whenever possible a King profile \citep{king66a, king66b} was fitted in the colour-magnitude (CM) filtered RDP, although such profiles are more effective for older clusters.} 



This paper is organized as follows: in Sec.\,\ref{clusters} we describe literature main aspects of the stellar clusters W\,31-CL, BDS\,113 and SGR\,$1806-20$. In Sec.\,\ref{data} the data sets are described. Sec. \ref{method} explains the analysis method and shows the results for the clusters. In Sec.\,\ref{discussion} we discuss them and in Sec.\,\ref{conclusions} the conclusions are given.

\section{Stellar Clusters in the W31 Complex}
\label{clusters}

\begin{figure*}
    \centering
    \includegraphics[width=0.45\textwidth]{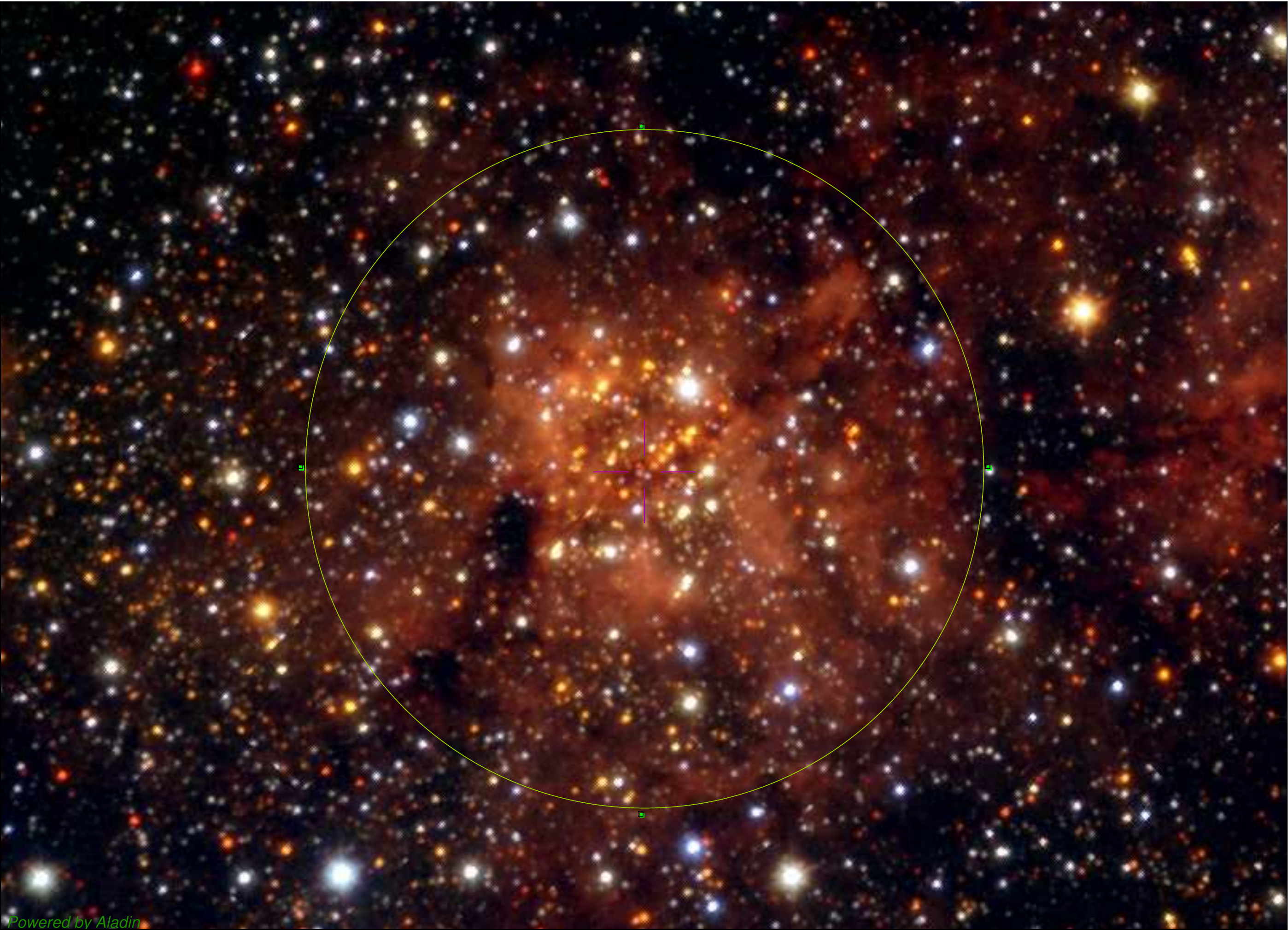}
    \includegraphics[width=0.45\textwidth]{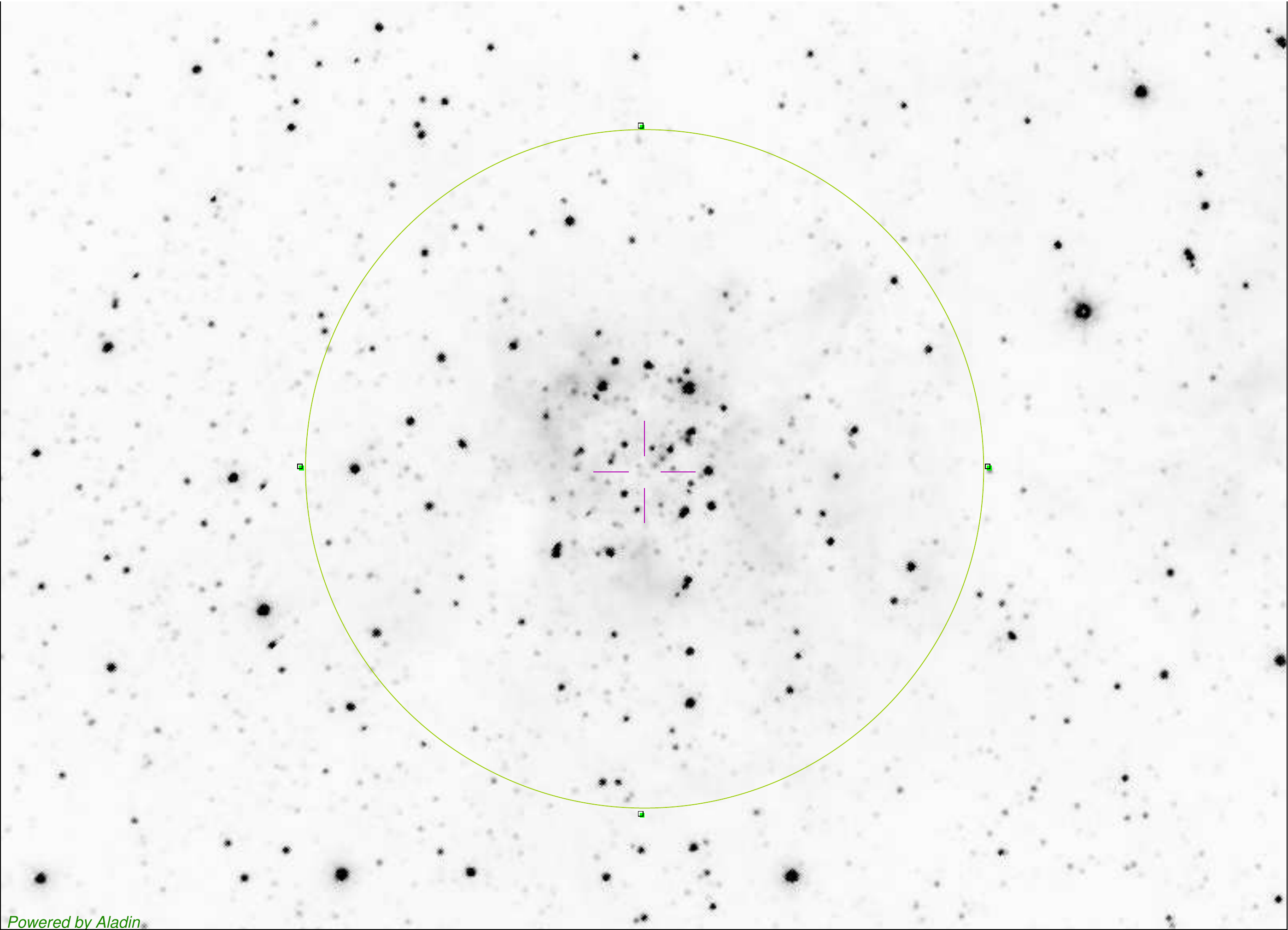}\\
    \includegraphics[width=0.45\textwidth]{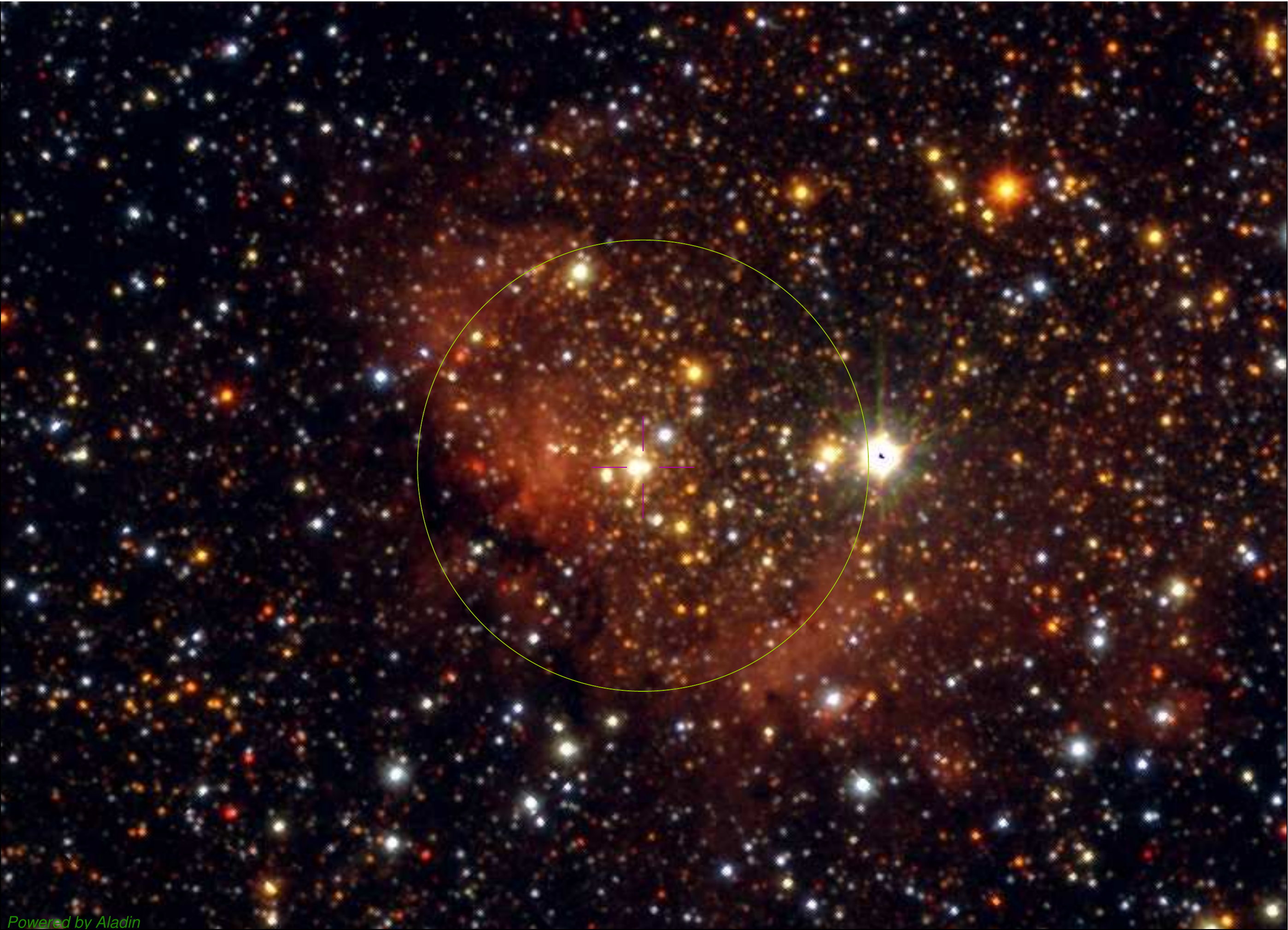}
    \includegraphics[width=0.45\textwidth]{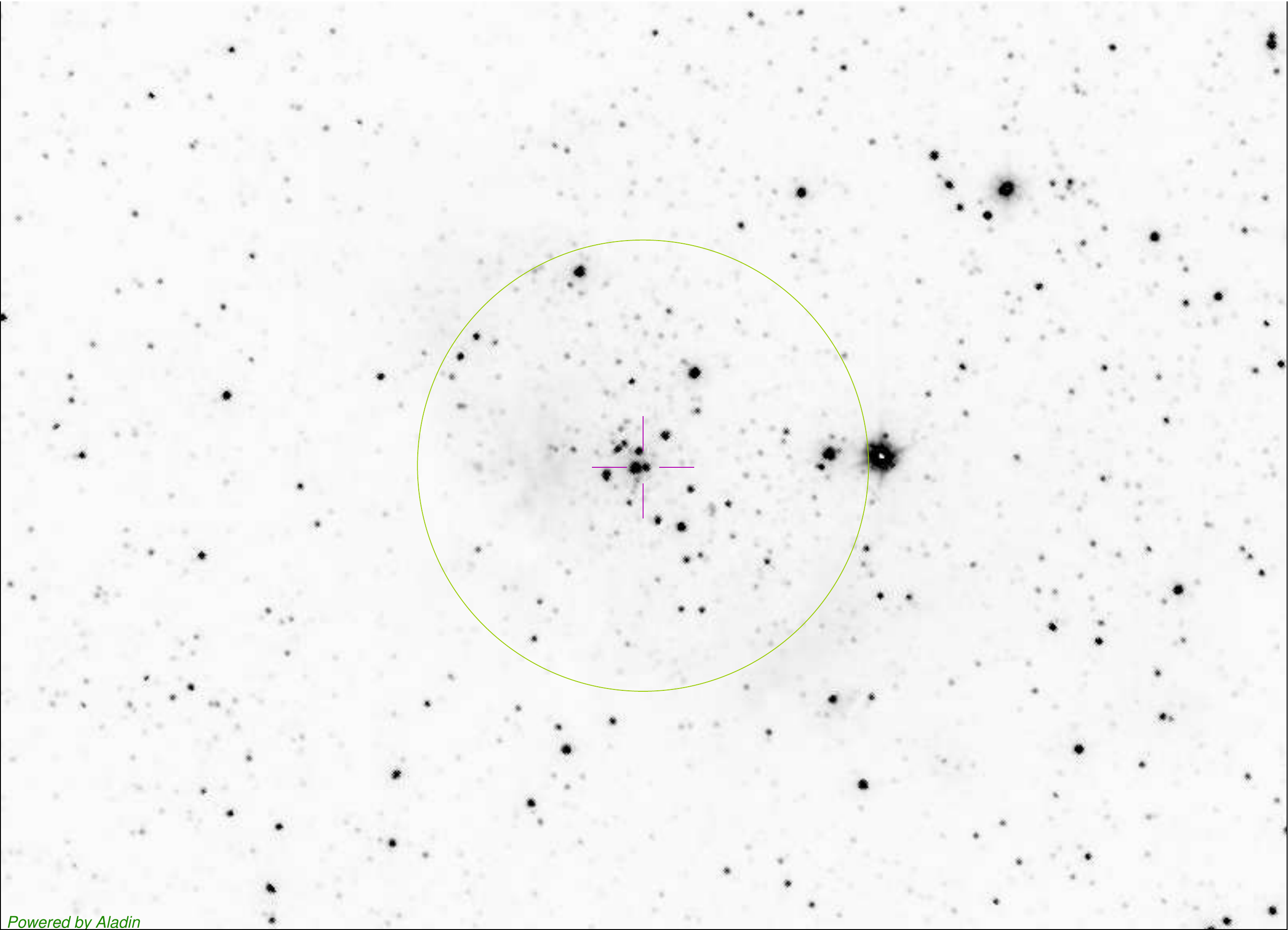}\\
    \includegraphics[width=0.45\textwidth]{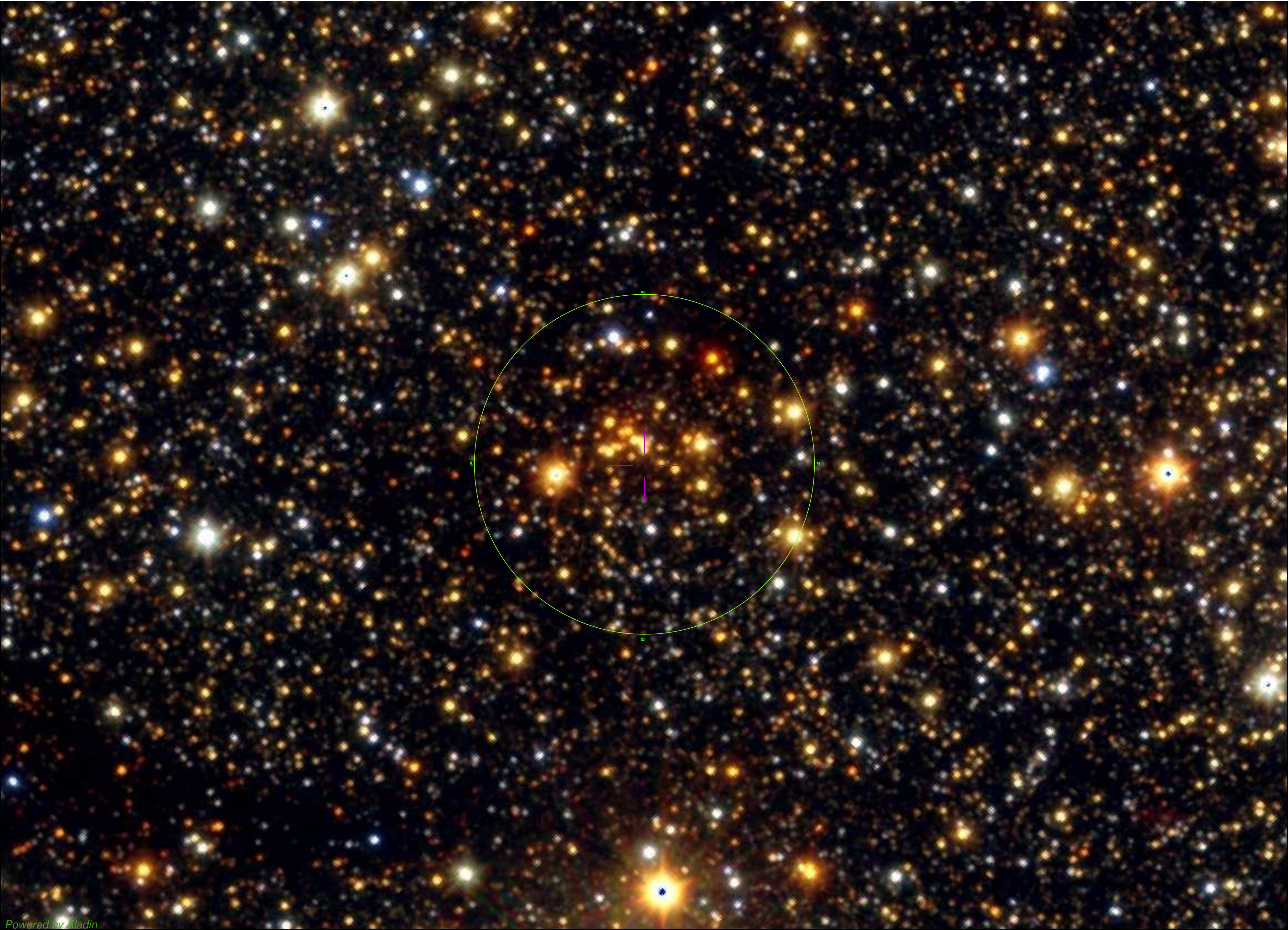}
    \includegraphics[width=0.45\textwidth]{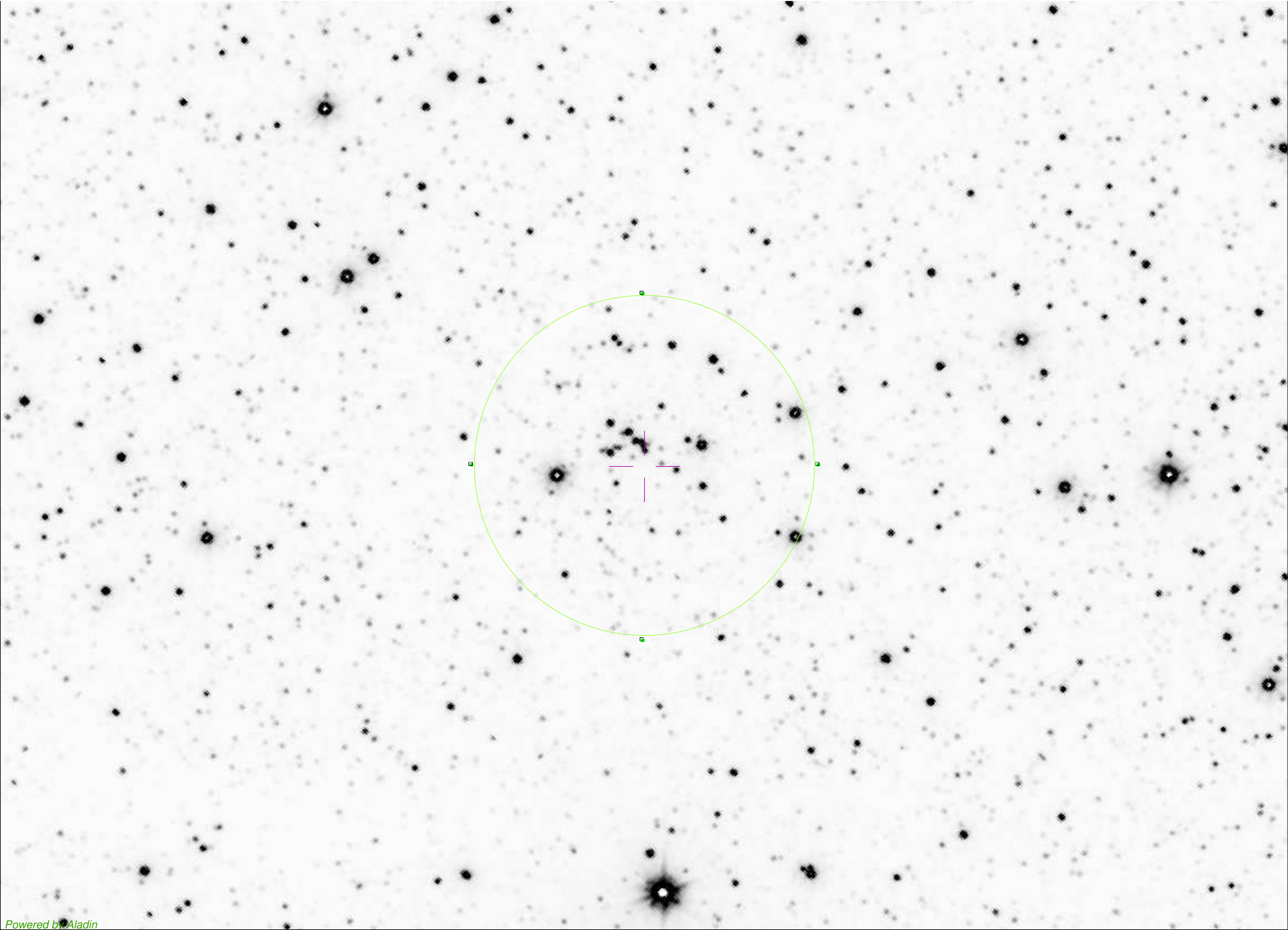}
    \caption{Three band (JHK$_s$) colour VVV image (left) and $K_S$-band (right) for the 3 clusters projected on or near the W\,31 complex. The images are centred at the clusters and have $4.54' \times 3.28'$ each. The green circles {($R=1.2'$, $R=0.8'$ and $R=0.6'$, respectively from top to bottom)} indicate the estimated cluster area. {North is up and east is left.}}
    \label{fig:clusters}
\end{figure*}

The W\,31 complex in Sagittarius was first observed by \citet{westerhout58} in radio wavelengths. Presently it appears to comprise at least four H{\sc ii} regions, two UCH{\sc ii}, four stellar clusters and one SGR. {Most of these objects are available within the boundaries of the VVV Survey. Table \ref{tbl:clusters} shows a summary of the clusters available in the survey and the related objects.} { Figure\,\ref{fig:clusters} we present the VVV three band (JHK$_S$) and $K_S$ images of the three stellar clusters available in the VVV survey: W\,31-CL, BDS\,113 and SGR\,$1806-20$.}

\begin{table*}
\caption{Clusters in the study: names, equatorial and Galactic coordinates, estimated radius, associated H{\sc ii} and UCH{\sc ii} regions and other associated objects.}
\centering
\scriptsize
\begin{tabular}{c c c c c c c c}
\hline
Clusters & $\alpha$ & $\delta$ & $l$ & $b$ & R & H{\sc ii}/UCH{\sc ii} regions & Other objects\\
 & (J2000) & (J2000) & ($^{\circ}$) & ($^{\circ}$) & { (')} &   \\
\hline
W\,31-CL &{\rm 18h09m27s} & {\rm $-20^{\circ}$19'30''} & 10.16 & $-0.36$ & 1.2 & G\,$10.2-0.3$, G\,$10.15-0.34$ & \\
BDS\,113 & {\rm 18h08m59s} &{\rm $-20^{\circ}$05'08''} & 10.32 & $-0.15$ & 0.8 & G\,$10.3-0.1$, G\,$10.3-0.15$ & CN\,148, IRAS\,$18060-2005$ \\
SGR\,$1806-20$ & {\rm 18h08m39s} & {\rm$-20^{\circ}$24'33''} & 10.00 & $-0.24$& 0.6 & G\,$10.0-0.3$ & AX\,$1805.7-2025$\\
\hline\\
\end{tabular}
\label{tbl:clusters}
\end{table*}

The main cluster component in the complex is W\,31-CL. 
It is associated with the H{\sc ii} region G\,$10.2-0.3$ and the UCH{\sc ii} region G\,$10.15-0.34$. {\citet{blum01} studied the stellar content and emphasised the presence of several O stars and young stellar objects (YSOs).} Based on the four O stars with $K$-band spectra, the authors determined a mean extinction of $A_V=15.5\pm1.7$\,mag. By assuming them as main sequence (MS), they derived a distance of $3.4$\,kpc and an age from the ZAMS to 1\,Myr.

BDS\,113 \citep{bica03} is the W\,31 complex component associated with the H{\sc ii} region G\,$10.3-0.1$, the UCH{\sc ii} G\,$10.3-0.15$, and the IRAS source IRAS\,$18060-2005$.
{\citet{dewangan15} analysed individual stars and continuum radio emission in the region of the mid infrared bubble CN\,148. They determined a mean cluster extinction of $A_V\sim14$\,mag and adopted a distance of $2.2$\,kpc, based on their references. They support triggered star formation, as a consequence of the expansion of the H{\sc ii} region.}

The most intriguing object in the set is the stellar cluster associated with SGR\,$1806-20$ \citep{fuchs99}. Besides the H{\sc ii} region G\,$10.0-0.3$, it is associated with an X-ray counterpart AX\,$1805.7-2025$ \citep{murakami94}. These authors also support the hypothesis that SGRs are related to neutron stars. \citet{figer05} studied the stellar content of the cluster SGR\,$1806-20$ and concluded that it has 3 Wolf-Rayet (WR) stars and one OB supergiant. {The presence of one LBV star is supported by \citet{eikenberry04}, and \citet{figer04} argue that it is a binary system. The stellar cluster is deeply absorbed by dust, internally and/or in the foreground, as revealed by its mean extinction of $A_V\sim30$\,mag \citep{corbel04}.}

The SGR\,$1806-20$ distance determination is an open point in the literature \citep{bibby08, corbel97, corbel04}. Some authors try to associate the SGR\,$1806-20$ with the W\,31 complex, but the results are not conclusive. \citet{bibby08} determined a distance of $8.7_{-1.5}^{+1.8}$\,kpc, while \citet{corbel04} obtained $15.1_{-1.3}^{+1.5}$\,kpc for the SGR  and G\,$10.3-0.1$, but $4.5\pm0.6$\,kpc for G\,$10.2-0.3$ and G\,$10.6-0.4$. {Solving this puzzle is a goal of the present study.}

\begin{figure}[t]
\centering
\includegraphics[trim={4cm 1.5cm 6.5cm 9cm},clip, width =0.45\textwidth]{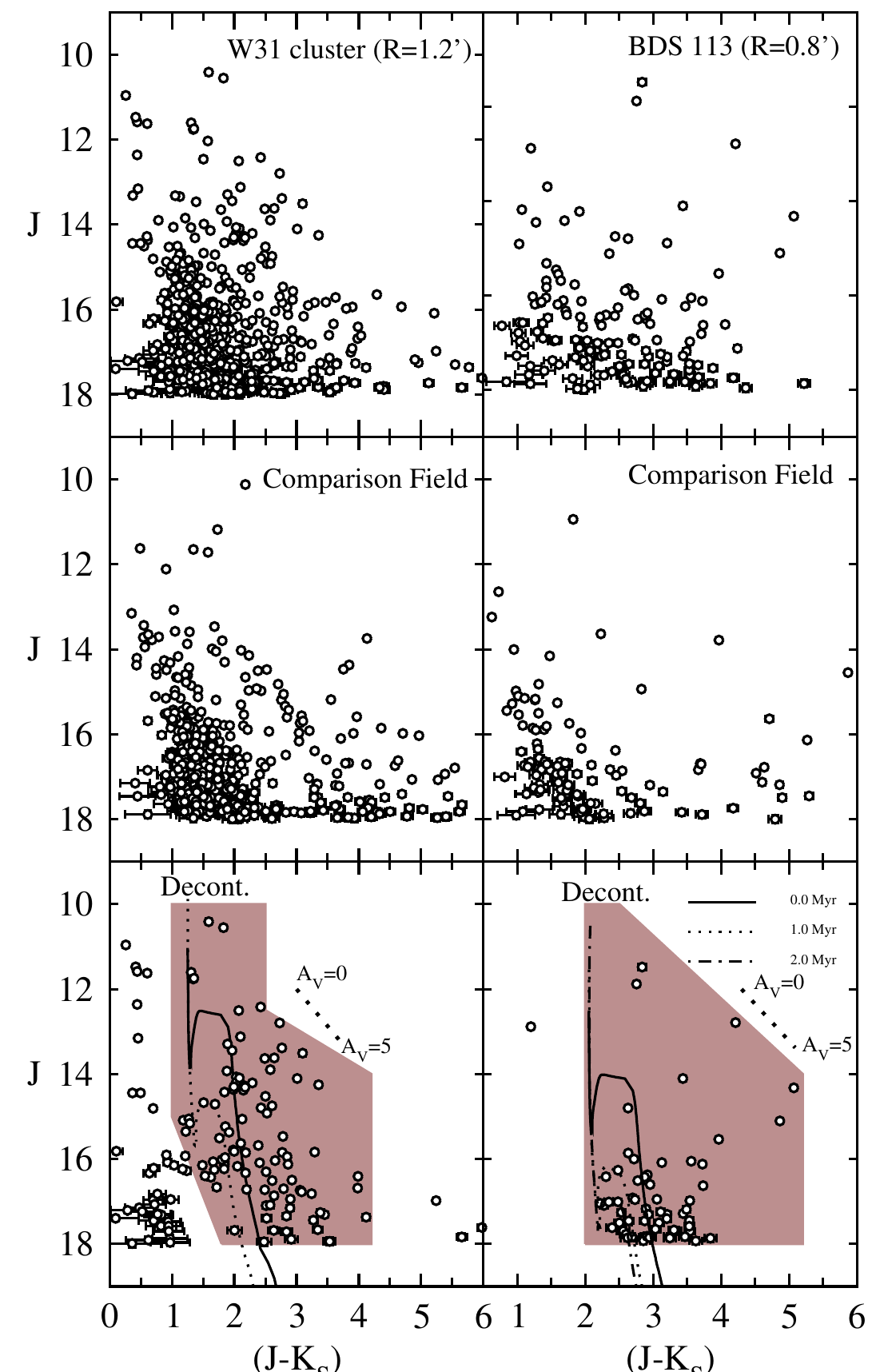}
\caption{Left panels: CMDs of W\,31-CL; right panels: CMDs of BDS\,113. The top panels show the stars in the observed cluster area. {Middle panels show the comparison field with the same projected area as the cluster}, and the bottom ones the field star decontaminated CMD. We overplot PARSEC isochrones.}
\label{fig:w31/bds113}
\end{figure}

\section{VVV photometry and methods}
\label{data}

{VVV is an ESO public survey that uses the VISTA telescope.}  It was designed to observe the bulge and part of the Milky Way disk in the NIR bands $Z$ ($0.87\mu$m), $Y$ ($1.02\mu$m), $J$ ($1.25\mu$m), $H$ ($1.64\mu$m) and $K_S$ ($2.14\mu$m). Among its goals is the study of RR-Lyrae stars in view of unveiling the central structure of the Galaxy {\citep{minniti10, saito12}}.

{The 4.1\,m VISTA telescope is located on Cerro Paranal, Chile. It has a Cassegrain focus and one instrument, the VIRCAM (Visual and InfraRed CAMera), described in details by \cite{dalton06}.}
 



The basic data reduction steps are performed by the Cambridge Astronomical Survey Unit (CASU) through the VISTA Data Flow System (VDFS), which provides positions, magnitudes and probable morphological type of the source. For details on the VVV photometry we refer to \citet{saito12}. {In the present study we used the aperture magnitudes provided by CASU\footnote{http://casu.ast.cam.ac.uk/vistasp/}.}

We employed the 2MASS photometry in the area to take into account saturation effects in the VVV photometry. The 2MASS photometry was obtained through the VizieR\footnote{http://vizier.u-strasbg.fr/viz-bin/VizieR} tool, in which we extract circular regions centered on the cluster coordinates. {The 2MASS magnitudes were converted into the VISTA Vegamag system according to the equations presented in \citet{soto13}. Stars brighter than $J$, $H$ and $K_S$ $\sim$11 mag in the VISTA Vegamag were replaced by the respective 2MASS counterpart, considering a maximum spatial separation of 0.5'' \citep{lima14}.}

\section{Star cluster analysis}
\label{method}
{
The VVV Survey deals with heavily contaminated fields, and so clusters in those regions require the stellar background decontamination. Our group have performed the same kind of analysis for the NGC\,6357 region \citep[and references therein]{lima14}, using CMDs, RDPs and field star decontamination to probe the features of ECs.}
Below we present a description of the methods used to analyze the VVV and 2MASS data of the selected stellar clusters.

\subsection{Field-star decontamination}
\label{fsd}


{Field-star (FS) decontamination is an essential tool to study stellar clusters projected on dense fields, like towards the disk and bulge. \citet{bonatto07} developed a FS decontamination algorithm based on the total number of FS in the cluster region. This method was developed for 2MASS $J$, $H$ and $K$ magnitudes, but we use an adapted version for the VVV photometry. Although the photometric and astrometric calibrations on the VVV data are performed via unsaturated 2MASS stars observed in each point, the VVV magnitudes are in the natural VISTA Vegamag system. {The cluster area is defined based on radial density profiles (RDPs) (see Sect.\ref{rdp}) and the comparison field is chosen according to the projected distribution of stars and the presence of dust around the cluster. For each cluster analysed in the present work we chose the ring, concentric to the cluster, that maximizes the FS decontamination efficiency and yet preserves the evolutionary tracks seen in the CMDs (Figs. \ref{fig:w31/bds113} and \ref{fig:blum_bik}).}

The method consists in dividing the stellar cluster CMD in a 3D grid with the magnitude $J$ and the colours $(J-H)$ and $(J-K_S)$ as the axes, with a uncertainty of $\sim 1\sigma$ in each direction. Each grid unit is called a cell and has typical dimensions of $\Delta J=2.0$\,mag and $\Delta(J-H)=\Delta(J-K_S)=0.5$\,mag.  These values are large enough to account for the photometric uncertainties and small enough to preserve the morphological properties of the evolutionary sequences. Subsequently, the expected number-density of FS in each cell is calculated based on the number of FS in  the comparison field with colours and magnitudes compatible with the ones of the cluster cells. The expected number of FS is randomly subtracted from each cell.}



Another FS decontamination procedure is the colour-magnitude (CM) filter. This method excludes the stars that are in different evolutionary sequences than the cluster and are outside to the filtered region in the CMD. We selected an area in the CMD in which the stars are more likely to be cluster members (brown areas in Figs. \ref{fig:w31/bds113}, \ref{fig:blum_bik}, \ref{fig:sgr06} and \ref{fig:sgr03}). A limitation of the CM filter is that it does not eliminate FS with similar colours and magnitudes as the cluster ones, although it is a very useful tool in the investigation of structural parameters with RDPs (Figures \ref{fig:rdp} and \ref{fig:rdp_sgr}) and the determination of mass functions. 

{In summary, (i) we build the FS decontaminated CMDs in order to determine physical parameters of the stellar clusters; and then (ii) we apply the CM filter to the ``raw'' data to create the filtered RDPs. These two procedures are essential to remove the effects of the dense stellar fields and allow the determination of the stellar and structural parameters of interest.} 

\subsection{Colour-magnitude diagrams}

\begin{figure}
    \centering
    \includegraphics[trim={2cm 1.5cm 11.5cm 9cm},clip, width=0.45\textwidth]{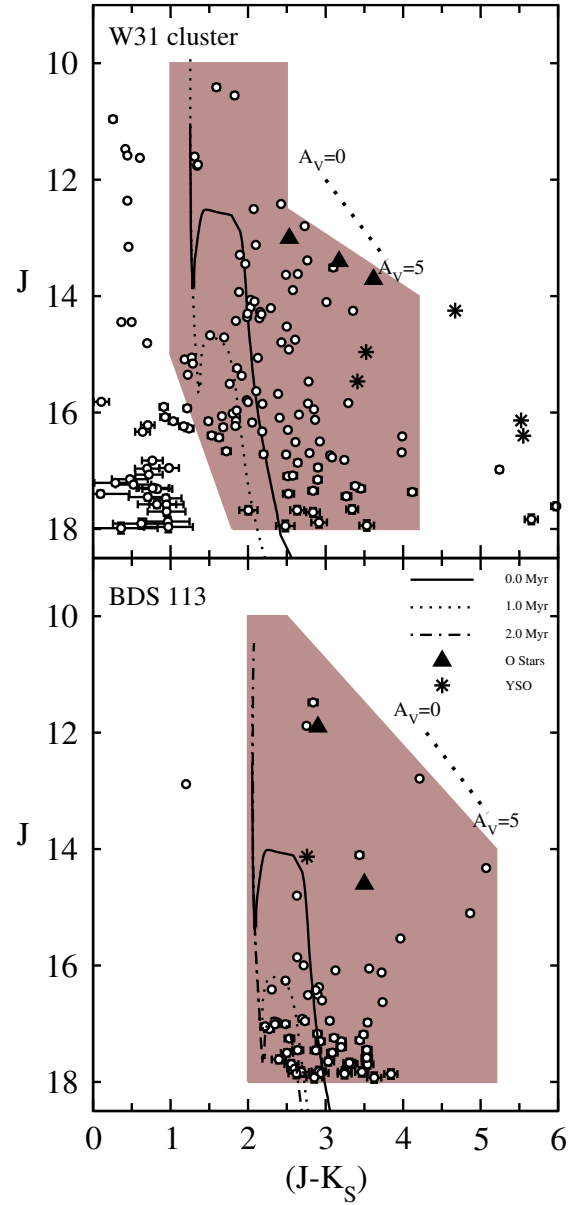}
    \caption{W\,31-CL and BDS\,113 CMDs: Same as bottom panels of Fig.\,\ref{fig:w31/bds113} with triangles and asterisms indicating the O stars and YSOs, respectively.  }
    \label{fig:blum_bik}
\end{figure}

The FS decontaminated CMDs are excellent tools to study stellar clusters. Through the isochrone fitting we can determine the age, distance and reddening of the cluster, which gives us a good idea of its physical scenario. The CMDs presented in Fig. \ref{fig:w31/bds113} show the following panels: (i) ``raw'' CMDs comprise all the stars in the cluster area; (ii) {comparison field CMDs with an area equal to cluster and the same range of colour and magnitude}; (iii) FS decontaminated CMDs show the  isochrone fitting, the CM filters, and the reddening vector. All the CMDs, including Figs. \ref{fig:blum_bik}, \ref{fig:sgr06} and \ref{fig:sgr03}, are plotted in the VVV bands $J\times (J-K_S)$.

The models used to derive the distance, age and reddening were the PAdova and TRieste Stellar Evolution Code (PARSEC\footnote{http://stev.oapd.inaf.it/cgi-bin/cmd}) from \citet{bressan12}. They provide theoretical modeling for both MS and PMS stars \citep{bonatto07,lima14}.

{The stellar parameters derived from the fitting are shown in Table \ref{tbl:parameters}, and include the value for the Galactocentric distance, $R_{GC}$, calculated by adopting the Sun's distance to the Galactic centre as $R_{\odot}=8.18$\,kpc \citep{abuter19} and the \citet{c89} exctition relations with $R_V=3.1$.}


\begin{table*}
\centering
\begin{tabular}{c c c c c c}
\hline
Clusters & Age\,(Myr) & $A_V$\,(mag) & Distance\,(kpc)& {R$_{GC}$\,(kpc)} \\
\hline
W\,31-CL & $0.5\pm0.5$ & $8.7\pm0.58$ & $4.5\pm1.09$ & $3.8\pm1.05$\\
BDS\,113 & $1\pm1$ & $13.3\pm0.58$ & $4.8\pm1.17$ & $3.5\pm1.09$ \\
SGR\,$1806-20$ & $10\pm4$ & $25.3\pm0.58$ & {$8.0\pm1.95$} & $1.4\pm0.49$\\
\hline
\end{tabular}
\caption{Fundamental parameters for the stellar clusters in this study. $R_{GC}$ was calculated using $R_{\odot}=8.18$\,kpc}
\label{tbl:parameters}
\end{table*}


\subsubsection{W\,31-CL}
W\,31-CL is the main infrared component of the star forming complex W\,31. Its decontaminated CMD (Figure \ref{fig:w31/bds113}, left boton panel), {for a selected radius of $1.2'$ and a comparison field of $R=4'-6'$,} shows two MS stars and a spread PMS indicating a wide range of masses for the YSOs in this region. This feature is common in ECs. 
{For the isochrone fitting we select a range of isochrones that basically reproduces the evolutionary sequences in the CMD, and determine the best fit. Errors correspond to parameter shifts with which the neighbouring isochrones marginally fit the data.

{\citet{blum01} analysed spectroscopically O-type and YSOs in W\,31. These stars basically lie along the reddening vector in the CMD of Fig.\,\ref{fig:blum_bik} (top panel). We must be dealing with early type stars with a wide range of absorption values, probably owing to embedding dust caps or cocoons.} Interestingly,  the possible reddening slope of these embedded massive stars suggests a shift from the normal reddening law, as represented by the reddening vector in the top panel of Fig.\,\ref{fig:blum_bik}.

W\,31-CL shows a considerable range of distances according to the literature, varying from $2.2$ to $4.5$\,kpc. However, the studies agree in the sense of a very young age, leading us to adopt two isochrones of zero age and 1\,Myr.} The parameters which led to the best {estimate} were: {$E(J-K_S)=1.5\pm0.10$, $E(B-V)=2.59\pm0.95$ and $A_V=8.72\pm0.58$. The observed distance modulus is $(m-M)_J=15.80\pm0.50$, and the absolute is $(m-M)_0=13.27\pm0.53$. We determined a distance of $d_\odot=4.5\pm1.09$\,kpc. }



\subsubsection{BDS\,113}
The BDS\,113 decontaminated CMD is shown in the right bottom panel of Fig.\,\ref{fig:w31/bds113}. The cluster radius considered for the FS decontamination was $R=0.8'$ {and a comparison field defined between $R=6'- 8'$}. It has no stars in the MS, so we based our fit only on the distribution of the PMS. There are two probable MS stars that show a $K_S$ excess, so they appear shifted to the right in the CMD. This CMD covers the same magnitude and colour range as the W\,31-CL one, which may indicate that both are physically connected. The PMS distribution in the CMD is characterised by a spread in colour. \citet{dewangan15} identified the YSOs in this region based on their infrared excess emission, observed due to the presence of natal envelopes or circumstellar disks. 

{\citet{bik05} identified two bright stars in the BDS\,113 region, one is a O8V-B2.5V and the other O5V-O6V. The CMD in the bottom panel of Fig.\,\ref{fig:blum_bik} presents these two stars with the YSO 2MASS\,J18085079-2005195. As seen in W\,31-CL the O stars lie along the reddening vector. As expected the YSO belongs to the PMS. }

{The fit of three PARSEC isochrones (0, 1 and 2\,Myr) was necessary to better describe the PMS distribution.}  The best {estimate} provided the following parameters: {$E(J-K_S)=2.30\pm0.10$, $E(B-V)= 3.97\pm0.95$ and $A_V=13.37\pm0.58$; the observed and absolute distance modules are $(m-M)_J=17.3\pm0.50$ and $(m-M)_0=13.42\pm0.53$, respectively. With these values we calculated a cluster distance of {$d_{\odot}=4.8\pm1.17$\,kpc}, which is comparable with the determination for W\,31-CL.} 

\subsubsection{SGR\,$1806-20$}
The literature has {already} placed the cluster in the direction of SGR\,$1806-20$ at about $d_{\odot}=15$\,kpc, but shorter distances were also considered (Sect. \ref{clusters}). {The present study with VVV photometric accuracy is a remarkable opportunity to shed light on this issue. The FS decontaminated CMD is shown in Figure \ref{fig:sgr03}. 
For the FS decontamination we employ $R=0.3'$, corresponding to the cluster high stellar density area (Sec.\,\ref{discussion}). The main difference from the previous two clusters is the high reddening, leading to a cluster sequence at $(J-K_S)\sim5$\,mag. The position of the LBV is indicated, when applicable. Its relationship, or not, with the cluster must be studied. In Sec.\,\ref{discussion} we also discuss the VVV counterpart of the cluster and field spectroscopic stars by \citet{figer05}.}



\begin{figure*}[ht]
    \centering
    \includegraphics[trim={1cm 0.5cm 1cm 8cm},clip,width=0.7\textwidth]{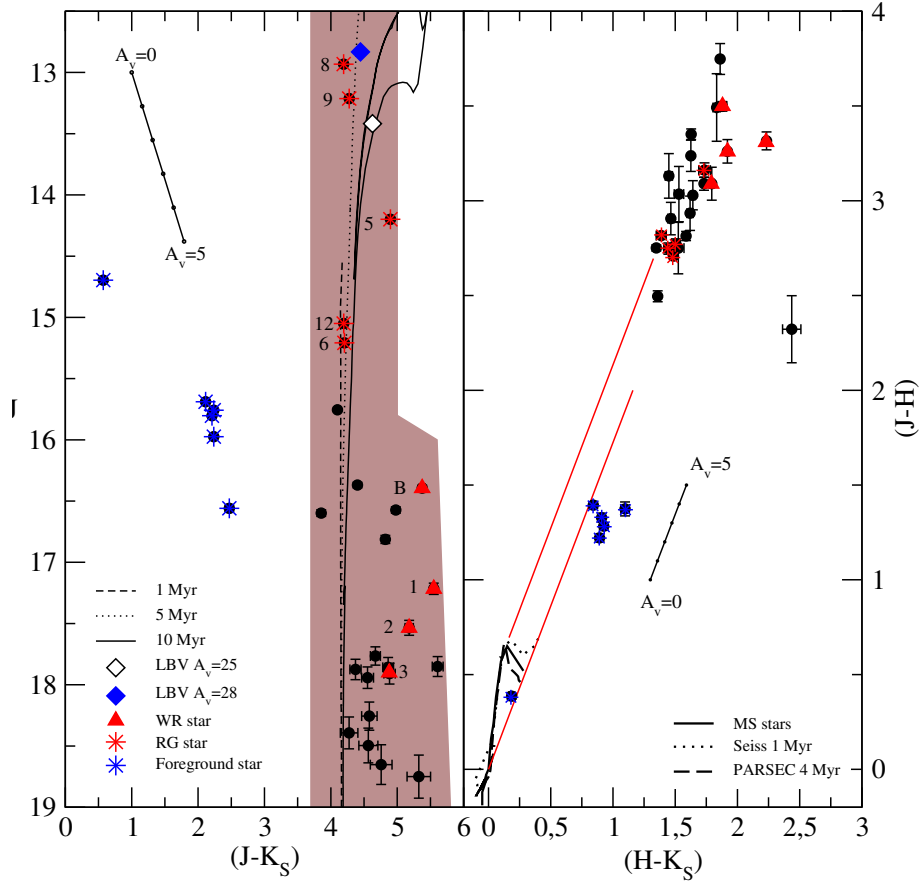}
    \caption{CMD (left panel) and CCD (right panel) {of the SGR\,1806-20 stars} in common between the present photometry and the spectroscopic sample of \citet{figer05}. Classes and labels of probable members and foreground stars are indicated by different symbols and labels in the left panel. Isochrone model is zero age from PARSEC. The stars are within $R=0.6$ (Fig.\,\ref{fig:clusters}).}
    \label{fig:sgr06}
\end{figure*}

\begin{figure*}[t]
    \centering
    \includegraphics[trim={1cm 0.5cm 1cm 8cm},clip,width=0.7\textwidth]{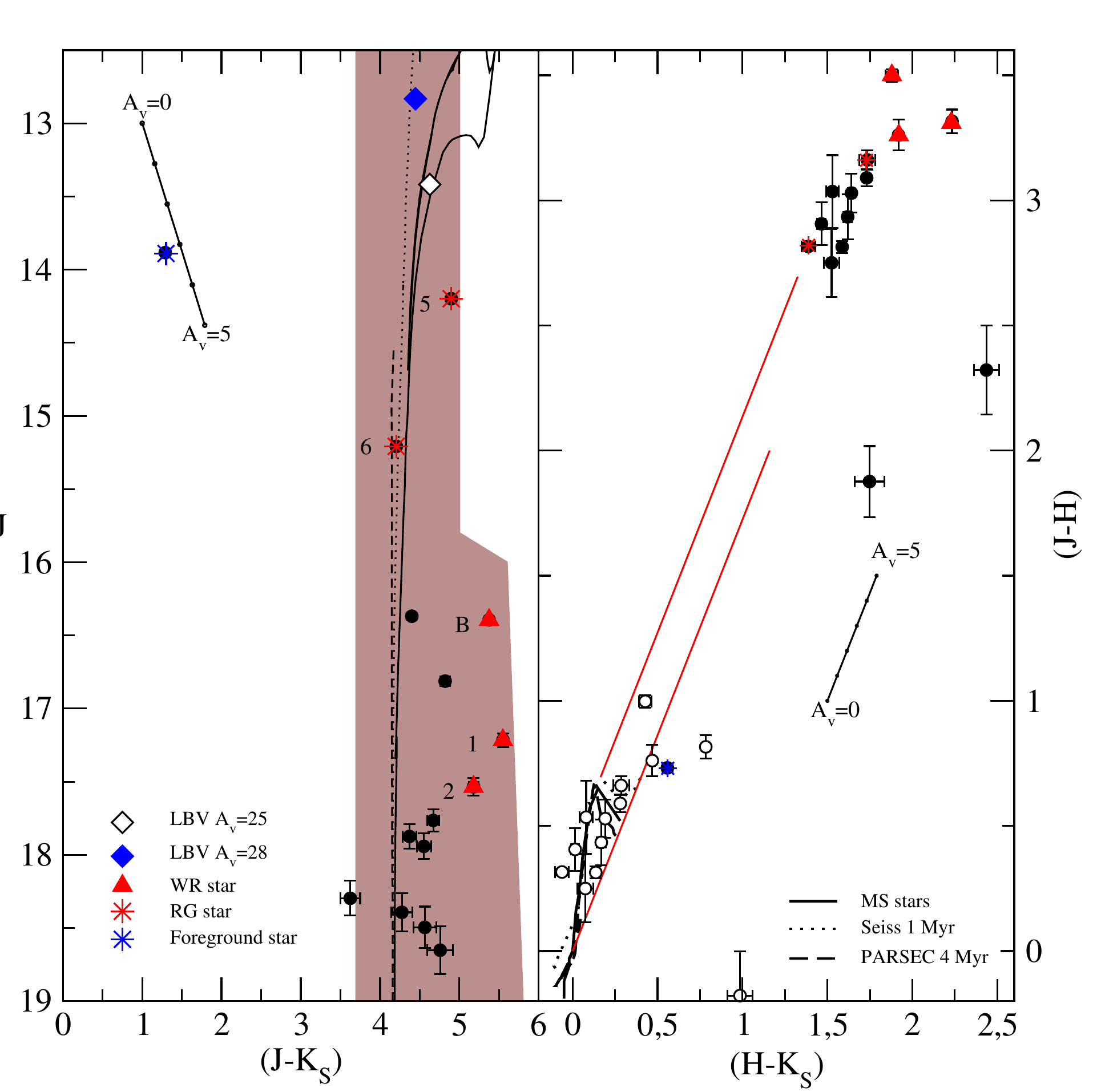}
    \caption{Same as Fig.\,\ref{fig:sgr06}, but for a present field decontamination photometry using the central high stellar density part of the cluster (R=$0.3$, see profile in Fig.\,\ref{fig:rdp_sgr} and central zoom image in Fig.\,\ref{fig:zoom}). The open circles in the CCD indicate the stars without the reddening effect.}
    \label{fig:sgr03}
\end{figure*}

\subsection{Radial density profiles}
\label{rdp}
The projected RDPs can be used to study the radial structure of a stellar cluster \citep[e.g.][]{bonatto07}. They are built by computing the density of stars in concentric rings. The centre is determined by visual inspection of a cluster image. We use the data before the FS decontamination and the CM filtered data, respectively. Usually the latter enables to separate the cluster from the field better than the former, as already demonstrated by \citet{bonattobica}.

The structural parameters are derived from the fit of a King-like profile \citep{king62} adapted to star counts, described by Equation \ref{eq:rdp}, where $\sigma(R)$, $\sigma_{bg}$ and $\sigma_{0}$ are the stellar, background and peak densities; $R$ and $R_c$ are the cluster and core radii.  We use a 2 parameter profile, which means that the $\sigma_{bg}$ is fixed and determined directly from the RDP.  These profiles also give an idea of the dynamical evolution of the cluster. 

\begin{equation}
\centering
{\rm \sigma(R)=\sigma_{bg}+\sigma_{0}/\left[1+(R/R_c)^2\right]}
\label{eq:rdp}
\end{equation}

In Figure \ref{fig:rdp} we show the RDPs for the ECs W\,31-CL and BDS\,113. The filled circles are the observed and the open the CM filtered RDPs. The 2 parameter King-like profile were fitted in the latter. Despite the small ages involved ($\sim 1$\,Myr) the two clusters fit well the King-like profile, which led to the following fitting parameters: $\sigma_{bg}=37.43\pm0.70$\,stars/arcmin$^2$, $\sigma_0=114.99$\,stars/arcmin$^2$ and $R_c=0.17\pm0.11'$ for W\,31-CL and $\sigma_{bg}=16.44\pm0.50$\,stars/arcmin$^2$, $\sigma_0=87.67\pm0.14$\,stars/arcmin$^2$ and $R_c=0.23\pm0.14'$ for BDS\,113. 
{We also built the RDP for SGR\,$1806-20$, presented in Figure \ref{fig:rdp_sgr}. The use of the CM filter clearly enhanced the cluster structure, and likewise the other two clusters we fitted a King profile. The background field density is $\sigma_{bg}=15.6\pm1.0$\,stars$/$arcmin$^2$, the peak density and core radius are $\sigma_0=106.31$\,stars/arcmin$^2$ and $R_c=0.17\pm0.07'$.
{The present clusters are young, therefore we do not expect relaxed systems. But interestingly, the three clusters can be fitted by a King-like profile. This is possibly related to formation and evolution of the clusters mimicking dynamically evolved profiles.}
\begin{figure}[!h]
\centering
\includegraphics[width=0.48\textwidth]{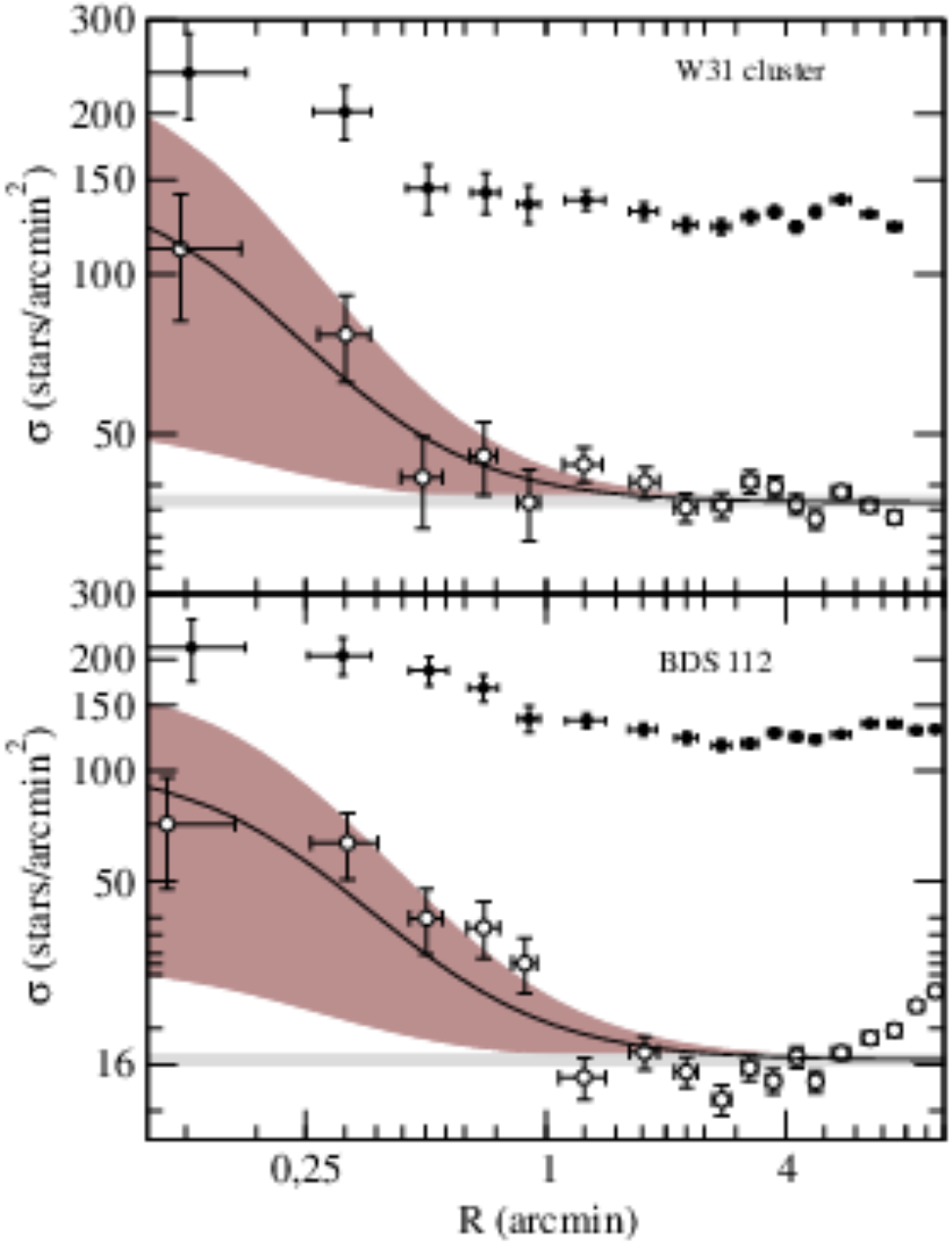}
\caption{Observed (filled circles) and {Colour-Magnitude} filtered (open circles) for W\,31-CL and BDS\,113 RDPs. A King-like profile was fitted (solid line). The 1$\sigma$ background level is the light-shaded region.}
\label{fig:rdp}
\end{figure}

{For the SGR\,$1806-20$ the cluster radius ($R<0.6'$) is first guessed from the cluster image and then optimized by the efficiency of the FS decontamination (Sec. \ref{disc_sgr}). This value is the one that determines the CMD extractions (Fig. \ref{fig:sgr06}).{The central overdensity ($R=0.3'$) is indicated in the RDP and determines the CMD extraction in Fig.\,\ref{fig:sgr03}.} In this case the input and output radii resulted the same.} The same procedure was adopted for the other two clusters.}
\begin{figure}[!h]
\centering
\includegraphics[width=0.48\textwidth]{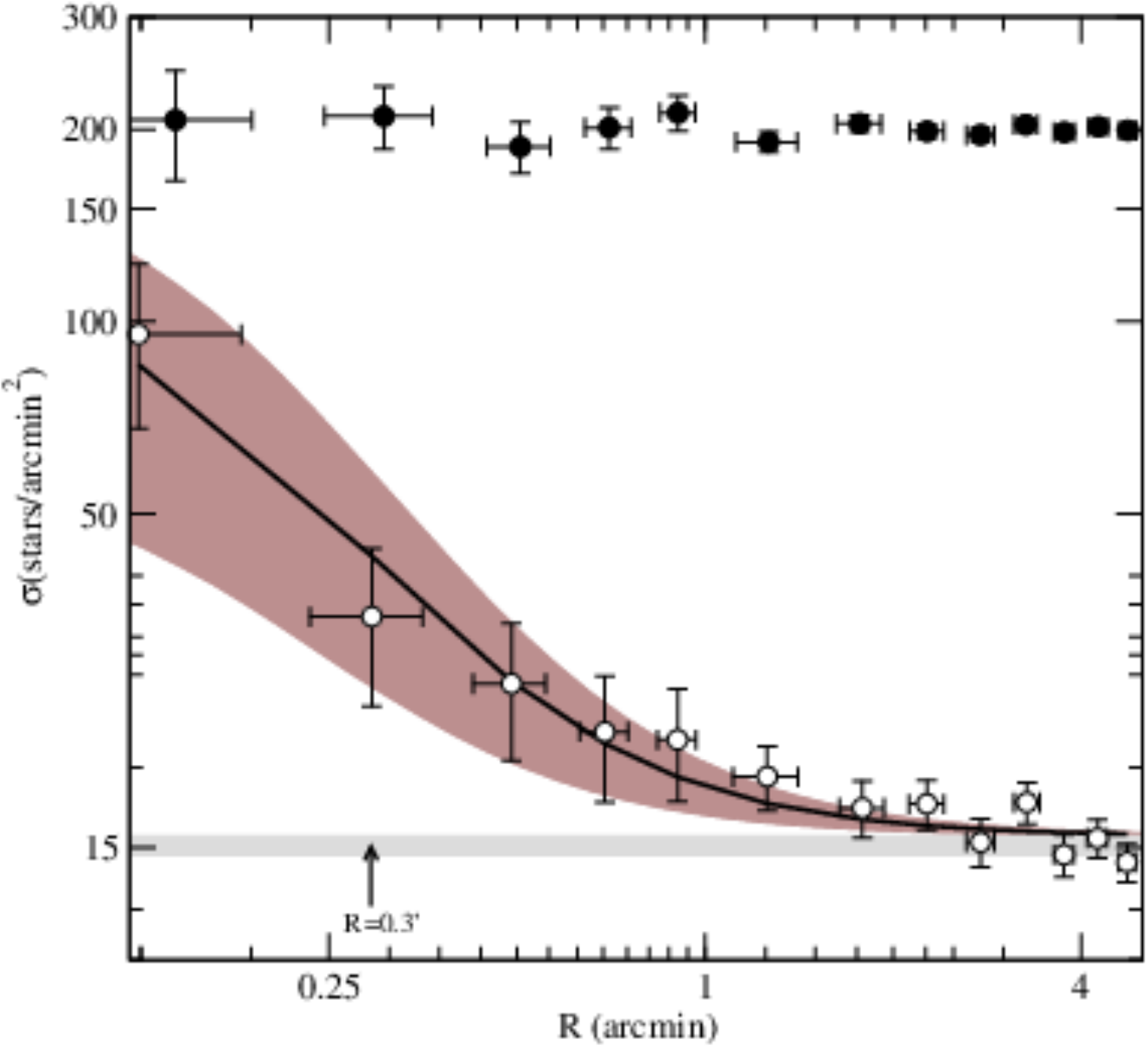}
\caption{Observed  and {Colour-Magnitude} filtered radial density profiles for SGR\,$1806-20$ in the top and bottom panels, respectively. Light-shaded region as in Figure \ref{fig:rdp}.}
\label{fig:rdp_sgr}
\end{figure}

\section{Discussion}
\label{discussion}
{
VVV allowed us to obtain for the three clusters deeper CMDs than available in the literature. Decontamination was fundamental to interpret them and derive astrophysical parameters. The present study of the ECs W\,31-CL and BDS\,113 showed that {they undoubtedly belong to the} W\,31 complex, and contain the ionizing sources of the cloud. The connection with the corresponding H{\sc ii} regions was suggested by \citet{kim02} and \citet{beuther11} in the radio domain. The distance determination for W\,31-CL, $d_{\odot}= 4.5$\,kpc, agrees with \citet{blum01}, however on a more straightforwardly basis by the present detection of the cluster PMS stars due to decontamination. We derive a distance of $d_{\odot}=4.8$\,kpc for BDS\,113, that is compatible with the W\,31 complex distance. In fact, the PMS stars are the main constraint for the isochrone fitting.  We emphasise that the present results were derived from stellar photometry, which is not dependent on the kinematic distance degeneracy problem \citep[e.g.][]{russeil}.}

{
We analysed the structure of these clusters by fitting RDPs with King-like profiles. The RDPs of W\,31-CL and BDS\,113 could {be basically described by this profile (Figure \ref{fig:rdp}). Such King-like solutions are certainly due to formation and early dynamical evolution effects that mimic much older evolved cluster RDPs. Crowding and dust absorption can also affect the counts in such profiles in ECs \citep{camargo15}.} It is noteworthy the importance of the decontamination procedures, in particular the CM filters \citep{bonatto07}. The observed profiles are essentially flat in such dense stellar fields, {as well as for clusters projected towards} the central disk and bulge.}

{\subsection{The nature of the cluster SGR\,$1806-20$}}
\label{disc_sgr}
{
In this section we discuss the properties of the star cluster SGR\,$1806-20$ to which the candidate LBV star and the Soft Gamma-Ray Repeater SGR\,$1806-20$ are projected within $R=0.6'$. In particular, we analyse colour-colour diagrams (CCD), which bring out the absorption univocally (right panels of Figs.\,\ref{fig:sgr06} and Fig.\,\ref{fig:sgr03}).

The cluster is deeply embedded with $A_V$=25\,mag, and this effect can be appreciated in the extreme red colours in the CMDs (left panels of Fig.\,\ref{fig:sgr06} and Fig.\,\ref{fig:sgr03}). This value is not at odds with $A_V$=28\,mag for the LBV \citep{figer05}. Fig.\,\ref{fig:sgr06} shows the CMD and CCD cross-identifications of the stars in the Keck spectroscopic analysis of \citet{figer05} and the present photometry. Part of these stars are listed in Table\,\ref{tbl:spec}, comparing possible luminous members with \citet{bibby08}. The CCD indicates probable members along the loci of clusters isochrone models. The symbols and labels allow to deduce the foreground stars that are bluer than the cluster sequences, and the red giants (RG) in the foreground put forward with spectroscopy by \citet{figer05}. In particular the plots indicate that the LBV is unique, with no rivals around in the CMD of Fig.\,\ref{fig:sgr06}, the bright neighboring stars in the CMD are foreground RGs. As to the luminous stellar content of the cluster, OI supergiants and WRs populate the base of the CMDs. WRs are redder than the cluster sequences in the CCD, which might suggest farther away. However dust caps can occur in them, so they are more probably members. All WRs basically match other luminous members in J, so there occurs no luminosity conflict in the cluster.}



{Fig.\,\ref{fig:zoom} shows 2 circles that correspond to the radii $R=0.6'$ delimitating the area of the cross-identification above with \citet{figer05}, and $R=0.3'$. The latter radius encompasses the denser central cluster area. The LBV star is indicated. The dense inner zone is ideal to apply our decontamination method. It has been employed to ECs in many previous studies \citep[e.g.][and references therein]{camargo12,camargo15}, as well as to the W\,31-CL and BDS\,113 in the present study. The field subtraction method \citep{bonatto07} ensures that we decontaminate the field stars from the bulge, disk and possible contamination from the W\,31 complex.}

\begin{table*}
\centering
\begin{tabular}{c c c c c c}
\hline
Stars & Spectral type  & $J_{VVV}$ & $(J-K_{S})_{VVV}$ & $K_{S\,VVV}$ & $K_{S\,GEM}$\\
(1) & (2) & (3) & (4) & (5) & (6)\\ 
\hline
\#1 & WC9d & 17.21 & 5.55 & 11.66 & 11.60\\
\#2 & WN6b & 17.53 & 5.18 & 12.35 & 12.16\\
\#3 & WN7 & 17.90 & 4.88 & 13.03 & 12.58 \\
\#4 & O9.5\,I & 17.94 & 4.55 & 13.38 & 11.92\\
\#7 & B0-B1\,I & 16.37 & 4.40 & 11.97 & 11.87\\
\#11 & B0\,I & 18.50 & 4.56 & 13.93 & 11.90\\
\#B & WC9d & 16.39 & 5.37 & 11.02 & 10.40\\
\#C & B1-B3\,I & 16.35 & 5.38 & 11.01 & 10.96\\
\#D & OB\,I & 17.85 & 4.86 & 13.00 & 11.06\\
\#A$^*$ & LBV & 13.66 & 4.92 & 8.74 & 9.26\\
\hline
\end{tabular}
\caption{Spectral and photometric information for the brighter stars in SGR\,$1806-20$. (1) and (2) Star designations and classification from \citet{figer05} and \citet{bibby08}; (3), (4), (5) $J$, $(J-K_S)$ and $K_S$ magnitudes and colours obtained from the present VVV photometry; (6) $K_S$ magnitude from \citet{bibby08}. *The LBV columns (2) and (6)  are from \citet{figer05} while columns (3) to (5) are from 2MASS database, replacing saturated VVV star.}
\label{tbl:spec}
\end{table*}

\begin{figure}[h]
    \includegraphics[trim={1cm 0.5cm 2cm 8cm},clip,width=0.47\textwidth]{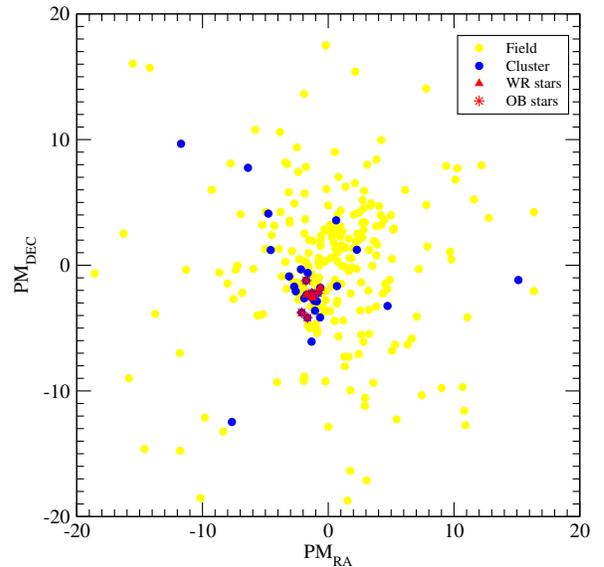}
    \caption{VVV PM for all the stars within the cluster radius ($R=0.6'$, yellow dots), and for the field decontaminated stars (blue dots) in SGR\,$1806-20$ (the same stars as in Fig.\,\ref{fig:sgr06}). RA PM corresponds to $\mu_{\alpha\cos\delta}$. {The WR and OB stars are indicated as red triangles and red asterisks.}}
    \label{fig:proper_motion}
\end{figure}

{The SGR\,$1806-20$ profile in Fig.\,\ref{fig:rdp_sgr}, shows that the cluster has a dense stellar core with $R=0.3'$. Its well populated, not a plain clump of bright stars, as at times reported in the early studies. Given its adequate structural properties we applied decontamination (left panels of Fig.\,\ref{fig:sgr06} and Fig.\,\ref{fig:sgr03}). Other young dense clusters with rich backgrounds were analysed with VVV in a similar way, e.g. Pismis\,24 and VVV CL 167 \citep{lima14}. The decontaminated CMD and CCD (Fig.\,\ref{fig:sgr06}) show that most foreground stars of \citet{figer05} were removed, either by the down-sized area and the cleaning with respect to the field. A PARSEC isochrone describes both the LBV evolved state and clusters bright stars below. We estimate from shifting solutions for neighbouring isochrones an age of $10\pm4$\,Myr and an heliocentric distance of $d_{\odot}=8.0\pm1.9$\,kpc. {The fitting parameters are: $E(J-K_S)=4.35\pm0.10$, $E(B-V)= 7.52\pm0.95$ and $A_V=25.29\pm0.58$; the observed and absolute distance modules are $(m-M)_J=21.86\pm0.50$ and $(m-M)_0=14.53\pm0.53$, respectively. They represent the best fitting solutions for the CMDs in Fig.\,\ref{fig:sgr06} and Fig.\,\ref{fig:sgr03}.}

The discovery and analysis of the new star clusters with VVV, both young and old, in the bulge or in the central disk have been recently boosted with the VVV Survey. Examples are the new globular cluster Minni\,22 and a series of new candidates \citep{minniti17}. Crowding in the bulge and central disk fields can make them virtually insconspicous in the VVV images, but they correspond to high stellar overdensities, in particular Minni\,22 is 30\,$\sigma$ above the background. In order to further stress the occurrence of a rather populous cluster in the direction of the soft-gamma ray source SGR\,$1806-20$, we use the first version of the {VVV Infrared Astrometric Catalogue \citep[VIRAC,][]{smith18}}. The proper motion (PM) components $\mu_{\alpha\cos\delta}$ and $\mu_{\delta}$ in mas\,yr$^{-1}$ for the field stars and the FS decontaminated stars of the SGR\,$1806-10$ are presented in Figure\,\ref{fig:proper_motion}. {The cluster stars show very similar PM values internally, specially the OB and WR stars, while the field has more spread values with and offset. This indicates that the stars in blue are physically connected. }VVV VIRAC efforts so far deal with magnitude ranges where the LBV is saturated, and consequently not present in the proper motion catalogues currently available for the region (e.g., VIRAC and Gaia; \citet{smith18,gaia}).



\begin{figure}[!t]
    \centering
    \includegraphics[width=0.45\textwidth]{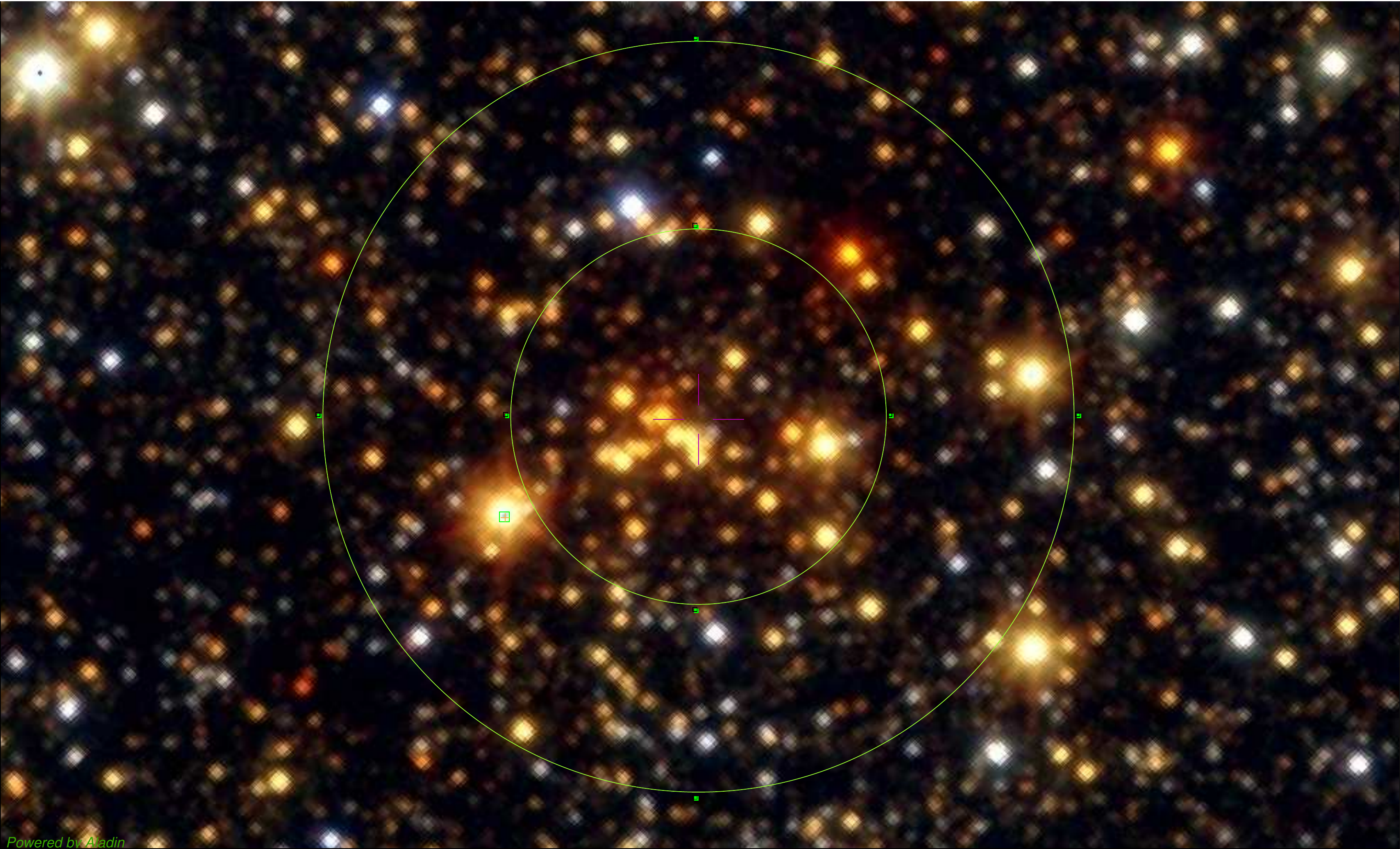}
    \caption{VVV image of SGR\,$1806-20$, zoomed with respect to Fig.\,\ref{fig:clusters}. The outer circle has $R=0.6'$ corresponding to the cross-identification area of Figer et al.'s photometry and the present photometry. The inner circle has $R=0.3'$ is ideal for the decontamination method. North to the top, east to the left, and field dimensions of $2.23'\times1.35'$.}
    \label{fig:zoom}
\end{figure}


Finally, we emphasise that in the past, there has been considerable spread in distance estimates of this cluster. The most recent one was by \citet{bibby08} leading to a cluster distance range of $7.2-10.4$\,kpc. The present value of {$8.0\pm1.95$\,kpc} is thus compatible, settling the distance issue. We conclude that the cluster SGR\,$1806-20$ is unrelated to the W\,31 complex, being almost a factor of $\approx1.8$ more distant.}

\section{Conclusions}
\label{conclusions}
{
We employed VVV photometry to study young clusters projected on or nearby the W\,31 star forming complex. We used CMDs and RDPs to derive cluster parameters, including structural ones. The W\,31 cluster and BDS\,113 belong to the complex at a distance of  $d_{\odot}=4.5-4.8$\,kpc. These two clusters appears to be the youngest generation in the complex with {ages around} $1$\,Myr. Note that the CMDs of W\,31-CL and BDS\,113 show clear PMS content as  in clusters like Pismis\,24 in the nearby complex NGC\,6357 \citep{lima14}. 

In the case of the SGR\,$1806-20$, we cross-identified the spectroscopic classification by \citet{figer05} with the present VVV/2MASS photometry in view of constraining the cluster parameters. We show how the stellar content, in terms of the LBV, WRs and O supergiants are mostly cluster members, and other stars belong to the foreground. The present cluster distance of $8.0$\,kpc is comparable with \citet{bibby08}. We conclude that the cluster SGR\,$1806-20$ is a factor of {1.7} more distant and significantly older than the W\,31 complex, and thus not physically related to it.}



\section*{Acknowledgments}
{We thank the anonymous referee for the comments that helped to improve the text.} This study was financed in part by the Coordena\c c\~ao de Aperfei\c coamento de Pessoal de N\'ivel Superior - Brasil (CAPES) - Finance Code 001, Conselho Nacional de Desenvolvimento Cient\'ifico e Tecnol\'ogico (CNPq) and Funda\c c\~ao de Amparo \`a pesquisa do Estado do RS (FAPERGS).
This publication makes use of data products from the Two Micron All Sky Survey, which is a joint project of the University of Massachusetts and the Infrared Processing and Analysis Center/California Institute of Technology, funded by the National Aeronautics and Space Administration and the National Science Foundation.
R.K.S. acknowledges support from CNPq/Brazil through projects 308968/2016-6 and 421687/2016-9.

\bibliographystyle{pasa-mnras}
\bibliography{refs}

\end{document}